\documentstyle[psfig,aps]{revtex}

\begin{document}

\pagestyle{plain}

\title{PROPERTIES OF PAIRING CORRELATIONS IN INFINITE NUCLEAR MATTER }

\author{\O.\ Elgar\o y}
\address{Department of Physics, University of Oslo, Norway}
\author{M.\ Hjorth-Jensen\footnote{Talk presented at Condensed Matter Theories 21,
Luso, Portugal, 21-27 september 1997. To appear in Condensed Matter Theories 
{\bf 13}.}}
\address{Nordita, Copenhagen, Denmark}

\maketitle

% old ugly as hell tex style for CMT21 proceedings, delete the % if you
% wish to embark on an excursion into the really dark old ages of computerized 
% mankind
%\magnification=\magstep1 
%\font\bigbfont=cmbx10 scaled\magstep1
%\font\bigifont=cmti10 scaled\magstep1
%\font\bigrfont=cmr10 scaled\magstep1
%\vsize = 23.5 truecm
%\hsize = 15.5 truecm
%\hoffset = .2truein
%\baselineskip = 14 truept
%\parskip = 3 truept
%\nopagenumbers

%\topinsert
%\vskip 3.2 truecm
%\endinsert

%\centerline{\bigbfont PROPERTIES OF PAIRING CORRELATIONS IN  }
%\vskip 3 truept
%\centerline{\bigbfont INFINITE NUCLEAR MATTER}
%\vskip 20 truept
%\centerline{\bigifont \O.\ Elgar\o y$^1$ and M.\ Hjorth-Jensen$^2$}
%\vskip 8 truept
%\centerline{\bigrfont $^1$Department of Physics, University of Oslo, Norway}
%\vskip 2 truept
%\centerline{\bigrfont $^2$Nordita, Copenhagen, Denmark}
%\vskip 1.8 truecm

\section{INTRODUCTION}
%\centerline{\bf 1.  INTRODUCTION}
%\vskip 12 truept

Recently, there has been renewed interest in the pairing problem in 
neutron matter and neutron-rich nuclei. The superfluid properties 
of neutron matter are of importance in the study of neutron stars 
[1], while pairing in neutron-rich systems is of relevance 
for the study of heavy nuclei close to the drip line [2]
and the light halo nuclei [3].  Much effort has gone into 
calculating the superfluid energy gap in dilute neutron matter 
[4-9].  Most of these studies have been carried 
out using pairing matrix elements given by the bare nucleon-nucleon 
(NN) interaction.  Even though it is a long time since  
Clark et al.\ [9] showed that density and spin-density 
fluctuations must be included in the pairing interaction, and there 
has been much progress in that direction recently [10,11], 
we will in this contribution focus on selected properties 
of the pairing problem in infinite neutron and nuclear matter
employing only the 
bare NN interaction.  

In this lowest-order approximation to the 
problem it has been found that all modern 
NN potentials give nearly identical results for the $^1S_0$ energy 
gap in dilute neutron matter.  One aim of this work is to 
explain how this can 
be understood directly from the measured properties of the free 
nucleon-nucleon (NN) interaction. This is discussed in Section 2.   

In Section 3 we discuss, still employing the bare NN interaction,
various properties of the pairing wave function.  
The pairing gap is determined by the attractive part
of the NN interaction. In the $^1S_0$ channel 
the potential is attractive for momenta $k \leq 1.74$ fm$^{-1}$
(or for interparticle distances $r \geq 0.57$ fm). However,
the nuclear situation is somewhat different from that of the classical BCS with
attractive potentials in the solid state, see e.g., the discussion in 
Ref.\ [12]. 
In the so-called weak coupling regime, where the 
interaction is weak and attractive,  a
gas of fermions may undergo a superconduncting (or superfluid) 
instability at low temperatures and a gas of Cooper pairs is formed.   
This gas of Cooper pairs will be surrounded by unpaired fermions and
the typical coherence length is large compared with the interparticle
spacing, and the bound pairs overlap. The latter behavior defines
also what we will mean with weak-coupling in this work. With weak-coupling
we will mean a regime where the coherence length is larger than the 
interparticle spacing.
In the so-called strong coupling limit, the formed bound pairs have only a small
overlap, the coherence length is small, and the bound pairs
can be treated as a gas of point bosons. One expects then the system
to undergo a Bose-Einstein condensation into a single quantum
state with total momentum $k=0$ [12].  
For the $^1S_0$ channel in nuclear physics we may actually expect to have 
two weak-coupling limits,
namely when the potential is weak and attractive for large interparticle
spacings and when the potential becomes repulsive at $r=0.57$ fm.
In these regimes, the potential has values of typically some few
MeV. One may also loosely speak of a strong-coupling limit 
where the NN potential is large 
and attractive. This takes place where the NN potential
reaches its maximum, with an absolute value of typically $\sim 100$ MeV, 
at roughly $\sim 1$ fm. 
These properties of the NN potential in the $^1S_0$ channel
and their connection with the wave function of the paired state
are discussed in Section 3. In that section we will argue, from the properties
of the wave function and the calculated coherence length, that 
fermion pairs in the $^1S_0$ wave in neutron and nuclear matter, will not
undergo the above-mentioned Bose-Einstein condensation, since, even though
the NN potential is large and attractive for certain Fermi momenta, the 
coherence length will always be larger than the interparticle spacing.

Concluding remarks and further
perspectives for pairing in nuclear systems are given in Section 4.
%\vskip 28 truept

%\centerline{\bf 2. PHASE-SHIFTS AND PAIRING GAP IN INFINITE MATTER }
%\vskip 12 truept
\section{PHASE-SHIFTS AND PAIRING GAP IN INFINITE MATTER }

The energy gap in infinite matter is obtained by solving the BCS equation 
for the gap function $\Delta(k)$.  
$$
      \Delta(k)=-{{1}\over {\pi}}\int_{0}^{\infty}dk'k'^2 
                 V(k,k'){{\Delta(k')}\over{E(k')}}, \eqno(1)
$$
where $V(k,k')$ is the bare momentum-space NN interaction in the 
$^1S_0$ channel, and $E(k)$ is the quasiparticle energy given by 
$E(k)=\sqrt{(\epsilon(k)-\epsilon(k_F))^2+\Delta(k)^2}$, where 
$\epsilon(k)$ is the single-particle energy of a neutron with 
momentum $k$, and $k_F$ is the Fermi momentum.  
Medium effects should 
be included in $\epsilon(k)$, but we will use free single-particle 
energies $\epsilon(k)=k^{2}/2m$, where $m$ is the neutron rest mass,  
to avoid unnecessary complications.  The omission of such medium
effects  is also in line with 
our omission of screening contributions.   
Anyway, at the densities 
considered here,  Brueckner-type calculations [7] 
indicate that in-medium single-particle energies do not 
differ much from the free ones. 
The energy gap is defined as $\Delta_F\equiv \Delta(k_F)$.  
Eq.\ (1) can be 
solved by various techniques, some of which are described in 
Refs.\ [7,8].  
In Fig.\ 1 we show 
the results for $\Delta_F$ obtained with the CD-Bonn potential (full line) 
[13],  
the Nijmegen I and Nimegen II potentials (dotted line and dashed line, 
respectively) [14]. 
The results are virtually identical, with the maximum value 
of the gap varying from 2.98 MeV for the Nijmegen I potential to 3.05 MeV 
for the Nijmegen II potential.  The same insensitivity of the results to 
the choice of NN interaction was found in Refs.\ 
[4,7].
We will now discuss how these results can be understood 
from the properties of the NN interaction in the $^1S_0$ channel.

A characteristic feature of $^1S_0$ NN scattering is the large, negative 
scattering length, indicating the presence of 
of a nearly bound state at zero scattering energy.  Near a bound state, 
where the NN $T$-matrix has a pole, it can be written in separable form, 
and this implies that the NN interaction itself to a good approximation is 
rank-one separable near this pole [15].   
Thus at low energies we can write 
$$
       V(k,k')=\lambda v(k)v(k'), \eqno(2)
$$
where $\lambda$ is a constant.  Then it is easily seen from 
Eq.\ (1) that the gap function can be written as $\Delta_F v(k)$, 
where $\Delta_F$ is the energy gap.  Inserting this form of 
$\Delta(k)$ into Eq.\ (1) one easily obtains 
$$
      1=-{{1}\over{\pi}}\int_{0}^{\infty}dk'k'^2{{\lambda v^2(k')}\over{E(k')}}.
      \eqno(3)
$$
%\vfill\eject
%\topinsert
%\vskip 6.1 truecm
%\endinsert
%\noindent
%{\bf Figure 1.}$^1S_0$ energy gap in neutron matter with the CD-Bonn, 
%             Nijmegen I and Nijmegen II potentials. In addition we 
%             show the results obtained from phase shifts only, Eqs.\ (3)-(5), 
%             and the effective range approximation of Eq.\ (6).
%\vskip 28 truept
%\noindent
Numerically the integral on the right-hand side of this equation depends 
very weakly on the momentum structure of $\Delta(k)$, so in our 
calculations we could take $\Delta(k)\approx \Delta_F$ in $E(k)$.  
Then Eq.\ (3) shows that the energy gap $\Delta_F$ is 
determined by the diagonal elements $\lambda v^2(k)$ of the NN interaction.  
The crucial point is that in scattering theory it can be shown that 
the inverse scattering problem, that is, the determination of a 
two-particle potential from the knowledge of the phase shifts at all 
energies, is exactly, and uniquely, solvable for rank-one 
separable potentials [16].  Following the notation 
of Ref.\ [15] we have 
$$
       \lambda v^2(k)=-{{k^2+\kappa_B^2}\over{k^2}}
                       {{\sin \delta(k)}\over{k}}e^{-\alpha(k)}, \eqno(4)
$$
for an attractive potential with a bound state at energy $E=-\kappa_B^2$. 
In our case $\kappa_B=0$.    
Here $\delta(k)$ is the $^1S_0$ phase shift as a function of momentum $k$, 
while $\alpha(k)$ is given by a principle value integral: 
$$
       \alpha(k)={{1}\over{\pi}}{\rm P}\int_{-\infty}^{+\infty}dk'
                 {{\delta(k')}\over{k'-k}},    \eqno(5)
$$
where the phase shifts are extended to negative momenta through 
$\delta(-k)=-\delta(k)$.

 From this discussion we see that $\lambda v^2(k)$, and therefore also 
the energy gap $\Delta_F$, is completely determined by the $^1S_0$ 
phase shifts.  However, there are two obvious limitations on the 
practical validity of this statement.  First of all, the separable 
approximation can only be expected to be good at low energies, near the 
pole in the $T$-matrix.  Secondly, we see from Eq.\ (5) that 
knowledge of the phase shifts $\delta(k)$ at all energies is required.  
This is, of course, impossible, and most phase shift 
analyses stop at a laboratory energy $E_{\rm lab}=350$ MeV.  The 
$^1S_0$ phase shift changes sign from positive to negative at 
$E_{\rm lab}\approx 248.5$ MeV
However, at low values of $k_F$, knowledge 
of $v(k)$ up to this value of $k$ may actually be enough to determine 
the value of $\Delta_F$, as the integrand in Eq.\ (3) is 
strongly peaked around $k_F$.  

The input in our calculation is the $^1S_0$ phase shifts taken from  
the recent Nijmegen phase shift analysis [17]. 
We then evaluated $\lambda v^2(k)$ from Eqs. (4) and 
(5), using methods described in Ref.\ [18] to 
evaluate the principle value integral in Eq.\ (5). 
Finally, we evaluated the energy gap $\Delta_F$ for various values 
of $k_F$ by solving Eq.\ (3), which is an algebraic 
equation due to the approximation $\Delta(k)\approx \Delta_F$ in the 
energy denominator.
\begin{figure}[hbtp]
    % GNUPLOT: LaTeX picture with Postscript
\setlength{\unitlength}{0.1bp}
\special{!
%!PS-Adobe-2.0
%%Creator: gnuplot
%%DocumentFonts: Helvetica
%%BoundingBox: 50 50 770 554
%%Pages: (atend)
%%EndComments
/gnudict 40 dict def
gnudict begin
/Color false def
/Solid false def
/gnulinewidth 5.000 def
/vshift -33 def
/dl {10 mul} def
/hpt 31.5 def
/vpt 31.5 def
/M {moveto} bind def
/L {lineto} bind def
/R {rmoveto} bind def
/V {rlineto} bind def
/vpt2 vpt 2 mul def
/hpt2 hpt 2 mul def
/Lshow { currentpoint stroke M
  0 vshift R show } def
/Rshow { currentpoint stroke M
  dup stringwidth pop neg vshift R show } def
/Cshow { currentpoint stroke M
  dup stringwidth pop -2 div vshift R show } def
/DL { Color {setrgbcolor Solid {pop []} if 0 setdash }
 {pop pop pop Solid {pop []} if 0 setdash} ifelse } def
/BL { stroke gnulinewidth 2 mul setlinewidth } def
/AL { stroke gnulinewidth 2 div setlinewidth } def
/PL { stroke gnulinewidth setlinewidth } def
/LTb { BL [] 0 0 0 DL } def
/LTa { AL [1 dl 2 dl] 0 setdash 0 0 0 setrgbcolor } def
/LT0 { PL [] 0 1 0 DL } def
/LT1 { PL [4 dl 2 dl] 0 0 1 DL } def
/LT2 { PL [2 dl 3 dl] 1 0 0 DL } def
/LT3 { PL [1 dl 1.5 dl] 1 0 1 DL } def
/LT4 { PL [5 dl 2 dl 1 dl 2 dl] 0 1 1 DL } def
/LT5 { PL [4 dl 3 dl 1 dl 3 dl] 1 1 0 DL } def
/LT6 { PL [2 dl 2 dl 2 dl 4 dl] 0 0 0 DL } def
/LT7 { PL [2 dl 2 dl 2 dl 2 dl 2 dl 4 dl] 1 0.3 0 DL } def
/LT8 { PL [2 dl 2 dl 2 dl 2 dl 2 dl 2 dl 2 dl 4 dl] 0.5 0.5 0.5 DL } def
/P { stroke [] 0 setdash
  currentlinewidth 2 div sub M
  0 currentlinewidth V stroke } def
/D { stroke [] 0 setdash 2 copy vpt add M
  hpt neg vpt neg V hpt vpt neg V
  hpt vpt V hpt neg vpt V closepath stroke
  P } def
/A { stroke [] 0 setdash vpt sub M 0 vpt2 V
  currentpoint stroke M
  hpt neg vpt neg R hpt2 0 V stroke
  } def
/B { stroke [] 0 setdash 2 copy exch hpt sub exch vpt add M
  0 vpt2 neg V hpt2 0 V 0 vpt2 V
  hpt2 neg 0 V closepath stroke
  P } def
/C { stroke [] 0 setdash exch hpt sub exch vpt add M
  hpt2 vpt2 neg V currentpoint stroke M
  hpt2 neg 0 R hpt2 vpt2 V stroke } def
/T { stroke [] 0 setdash 2 copy vpt 1.12 mul add M
  hpt neg vpt -1.62 mul V
  hpt 2 mul 0 V
  hpt neg vpt 1.62 mul V closepath stroke
  P  } def
/S { 2 copy A C} def
end
}
\begin{picture}(3600,2160)(0,0)
\special{"
gnudict begin
gsave
50 50 translate
0.100 0.100 scale
0 setgray
/Helvetica findfont 100 scalefont setfont
newpath
-500.000000 -500.000000 translate
LTa
600 251 M
2817 0 V
600 251 M
0 1858 V
LTb
600 251 M
63 0 V
2754 0 R
-63 0 V
600 437 M
63 0 V
2754 0 R
-63 0 V
600 623 M
63 0 V
2754 0 R
-63 0 V
600 808 M
63 0 V
2754 0 R
-63 0 V
600 994 M
63 0 V
2754 0 R
-63 0 V
600 1180 M
63 0 V
2754 0 R
-63 0 V
600 1366 M
63 0 V
2754 0 R
-63 0 V
600 1552 M
63 0 V
2754 0 R
-63 0 V
600 1737 M
63 0 V
2754 0 R
-63 0 V
600 1923 M
63 0 V
2754 0 R
-63 0 V
600 2109 M
63 0 V
2754 0 R
-63 0 V
600 251 M
0 63 V
0 1795 R
0 -63 V
976 251 M
0 63 V
0 1795 R
0 -63 V
1351 251 M
0 63 V
0 1795 R
0 -63 V
1727 251 M
0 63 V
0 1795 R
0 -63 V
2102 251 M
0 63 V
0 1795 R
0 -63 V
2478 251 M
0 63 V
0 1795 R
0 -63 V
2854 251 M
0 63 V
0 1795 R
0 -63 V
3229 251 M
0 63 V
0 1795 R
0 -63 V
600 251 M
2817 0 V
0 1858 V
-2817 0 V
600 251 L
LT0
1814 1946 M
180 0 V
788 282 M
2 1 V
2 1 V
4 2 V
5 3 V
6 3 V
8 3 V
8 5 V
10 5 V
11 7 V
12 7 V
13 8 V
14 10 V
16 11 V
16 12 V
18 13 V
18 15 V
20 16 V
21 18 V
21 19 V
23 22 V
23 22 V
25 25 V
25 26 V
26 27 V
27 29 V
28 30 V
29 32 V
29 32 V
31 34 V
31 35 V
31 36 V
33 36 V
32 36 V
34 37 V
34 37 V
35 37 V
35 37 V
35 36 V
36 35 V
36 34 V
37 33 V
37 32 V
37 30 V
38 28 V
37 26 V
38 24 V
38 22 V
38 19 V
38 17 V
39 13 V
38 12 V
38 8 V
38 4 V
38 2 V
37 -2 V
38 -4 V
37 -8 V
37 -10 V
37 -14 V
36 -17 V
36 -19 V
35 -22 V
35 -25 V
35 -27 V
34 -28 V
34 -31 V
32 -32 V
33 -34 V
31 -36 V
31 -36 V
31 -37 V
29 -37 V
29 -36 V
28 -37 V
27 -37 V
26 -36 V
25 -35 V
25 -34 V
23 -32 V
23 -31 V
21 -29 V
21 -28 V
20 -26 V
18 -24 V
18 -22 V
16 -21 V
16 -19 V
14 -17 V
13 -15 V
12 -14 V
11 -12 V
10 -10 V
8 -10 V
8 -7 V
6 -7 V
5 -5 V
4 -4 V
2 -3 V
2 -1 V
LT1
1814 1846 M
180 0 V
788 281 M
2 1 V
2 1 V
4 2 V
5 2 V
6 3 V
8 4 V
8 4 V
10 5 V
11 6 V
12 8 V
13 8 V
14 9 V
16 10 V
16 12 V
18 13 V
18 15 V
20 16 V
21 18 V
21 19 V
23 21 V
23 22 V
25 24 V
25 26 V
26 27 V
27 28 V
28 30 V
29 31 V
29 32 V
31 34 V
31 34 V
31 35 V
33 36 V
32 36 V
34 37 V
34 36 V
35 36 V
35 36 V
35 36 V
36 35 V
36 34 V
37 32 V
37 31 V
37 30 V
38 28 V
37 25 V
38 24 V
38 21 V
38 19 V
38 16 V
39 13 V
38 10 V
38 7 V
38 5 V
38 1 V
37 -2 V
38 -5 V
37 -8 V
37 -12 V
37 -14 V
36 -17 V
36 -20 V
35 -22 V
35 -25 V
35 -27 V
34 -30 V
34 -31 V
32 -32 V
33 -34 V
31 -36 V
31 -36 V
31 -36 V
29 -37 V
29 -37 V
28 -36 V
27 -37 V
26 -35 V
25 -35 V
25 -33 V
23 -32 V
23 -30 V
21 -29 V
21 -27 V
20 -25 V
18 -24 V
18 -21 V
16 -20 V
16 -18 V
14 -17 V
13 -14 V
12 -14 V
11 -11 V
10 -10 V
8 -9 V
8 -7 V
6 -6 V
5 -5 V
4 -4 V
2 -2 V
2 -2 V
LT2
1814 1746 M
180 0 V
788 282 M
2 1 V
2 1 V
4 2 V
5 2 V
6 3 V
8 4 V
8 5 V
10 5 V
11 6 V
12 8 V
13 8 V
14 10 V
16 10 V
16 12 V
18 14 V
18 14 V
20 17 V
21 18 V
21 19 V
23 21 V
23 23 V
25 24 V
25 26 V
26 28 V
27 28 V
28 31 V
29 31 V
29 33 V
31 34 V
31 35 V
31 36 V
33 36 V
32 36 V
34 37 V
34 37 V
35 37 V
35 37 V
35 36 V
36 35 V
36 35 V
37 33 V
37 32 V
37 30 V
38 29 V
37 26 V
38 25 V
38 22 V
38 19 V
38 17 V
39 14 V
38 12 V
38 8 V
38 5 V
38 2 V
37 -1 V
38 -4 V
37 -8 V
37 -10 V
37 -14 V
36 -16 V
36 -19 V
35 -22 V
35 -25 V
35 -26 V
34 -29 V
34 -31 V
32 -32 V
33 -34 V
31 -35 V
31 -37 V
31 -36 V
29 -37 V
29 -37 V
28 -37 V
27 -37 V
26 -36 V
25 -36 V
25 -34 V
23 -32 V
23 -32 V
21 -29 V
21 -28 V
20 -26 V
18 -25 V
18 -22 V
16 -21 V
16 -19 V
14 -18 V
13 -15 V
12 -14 V
11 -12 V
10 -11 V
8 -10 V
8 -7 V
6 -7 V
5 -5 V
4 -4 V
2 -3 V
2 -1 V
LT3
1814 1646 M
180 0 V
788 274 M
2 0 V
2 0 V
4 1 V
5 0 V
6 1 V
8 2 V
8 2 V
10 2 V
11 3 V
12 4 V
13 5 V
14 5 V
16 7 V
16 8 V
18 9 V
18 11 V
20 12 V
21 14 V
21 16 V
23 17 V
23 19 V
25 21 V
25 23 V
26 25 V
27 27 V
28 29 V
29 31 V
29 32 V
31 34 V
31 35 V
31 37 V
33 37 V
32 37 V
34 38 V
34 39 V
35 39 V
35 39 V
35 38 V
36 39 V
36 37 V
37 37 V
37 35 V
37 34 V
38 32 V
37 31 V
38 28 V
38 27 V
38 23 V
38 22 V
39 19 V
38 16 V
38 13 V
38 9 V
38 7 V
37 3 V
38 0 V
37 -3 V
37 -6 V
37 -10 V
36 -13 V
36 -16 V
35 -19 V
35 -22 V
35 -25 V
34 -27 V
34 -29 V
32 -32 V
33 -33 V
31 -35 V
31 -37 V
31 -37 V
29 -38 V
29 -38 V
28 -38 V
27 -38 V
26 -37 V
25 -36 V
25 -36 V
23 -34 V
23 -33 V
21 -32 V
21 -31 V
20 -29 V
18 -27 V
18 -26 V
16 -25 V
16 -22 V
14 -21 V
13 -19 V
12 -18 V
11 -16 V
10 -14 V
8 -12 V
8 -11 V
6 -9 V
5 -7 V
4 -5 V
2 -4 V
2 -2 V
LT4
1814 1546 M
180 0 V
788 276 M
2 1 V
2 1 V
4 1 V
5 2 V
6 2 V
8 4 V
8 3 V
10 5 V
11 5 V
12 7 V
13 7 V
14 9 V
16 9 V
16 11 V
18 13 V
18 14 V
20 15 V
21 17 V
21 19 V
23 20 V
23 22 V
25 24 V
25 25 V
26 28 V
27 28 V
28 30 V
29 32 V
29 33 V
31 34 V
31 36 V
31 37 V
33 37 V
32 39 V
34 40 V
34 40 V
35 41 V
35 41 V
35 42 V
36 42 V
36 42 V
37 42 V
37 42 V
37 41 V
38 40 V
37 40 V
38 39 V
38 38 V
38 37 V
38 37 V
39 35 V
38 33 V
38 33 V
38 30 V
38 30 V
37 27 V
38 27 V
37 24 V
37 23 V
37 22 V
36 20 V
36 18 V
35 17 V
35 15 V
35 14 V
34 12 V
34 11 V
32 10 V
33 8 V
31 7 V
31 6 V
31 5 V
29 4 V
29 2 V
28 2 V
27 1 V
26 1 V
25 -1 V
25 0 V
23 -2 V
23 -2 V
21 -2 V
21 -3 V
20 -3 V
18 -3 V
18 -3 V
16 -3 V
16 -3 V
14 -4 V
13 -3 V
12 -3 V
11 -3 V
10 -2 V
8 -3 V
8 -2 V
6 -1 V
5 -2 V
4 -1 V
2 -1 V
2 0 V
stroke
grestore
end
showpage
}
\put(1754,1546){\makebox(0,0)[r]{Effective range}}
\put(1754,1646){\makebox(0,0)[r]{Phase shifts}}
\put(1754,1746){\makebox(0,0)[r]{Nijmegen II}}
\put(1754,1846){\makebox(0,0)[r]{Nijmegen I}}
\put(1754,1946){\makebox(0,0)[r]{CD-Bonn}}
\put(2008,21){\makebox(0,0){$k_F$ fm$^{-1}$}}
\put(100,1180){%
\special{ps: gsave currentpoint currentpoint translate
270 rotate neg exch neg exch translate}%
\makebox(0,0)[b]{\shortstack{Pairing gap $\Delta (k_F)$ (MeV)}}%
\special{ps: currentpoint grestore moveto}%
}
\put(3229,151){\makebox(0,0){1.4}}
\put(2854,151){\makebox(0,0){1.2}}
\put(2478,151){\makebox(0,0){1}}
\put(2102,151){\makebox(0,0){0.8}}
\put(1727,151){\makebox(0,0){0.6}}
\put(1351,151){\makebox(0,0){0.4}}
\put(976,151){\makebox(0,0){0.2}}
\put(600,151){\makebox(0,0){0}}
\put(540,2109){\makebox(0,0)[r]{5}}
\put(540,1923){\makebox(0,0)[r]{4.5}}
\put(540,1737){\makebox(0,0)[r]{4}}
\put(540,1552){\makebox(0,0)[r]{3.5}}
\put(540,1366){\makebox(0,0)[r]{3}}
\put(540,1180){\makebox(0,0)[r]{2.5}}
\put(540,994){\makebox(0,0)[r]{2}}
\put(540,808){\makebox(0,0)[r]{1.5}}
\put(540,623){\makebox(0,0)[r]{1}}
\put(540,437){\makebox(0,0)[r]{0.5}}
\put(540,251){\makebox(0,0)[r]{0}}
\end{picture}
    \caption{$^1S_0$ energy gap in neutron matter with the CD-Bonn, 
             Nijmegen I and Nijmegen II potentials. In addition we 
             show the results obtained from phase shifts only, Eqs.\ (3)-(5), 
             and the effective range approximation of Eq.\ (6).}
\end{figure}
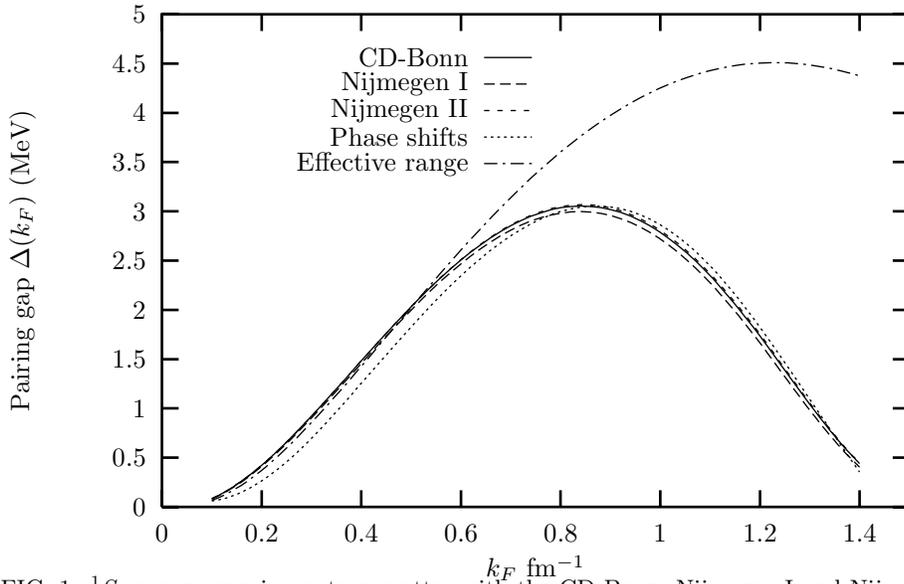
The resulting energy gap is plotted in Fig.\ 
1 (full line).
As the reader can see, the agreement 
between the direct calculation from the phase shifts and the CD-Bonn and
Nijmegen 
calculation of $\Delta_F$ is, to say the least, satisfying, even 
at densities as high as $k_F=1.4\;{\rm fm}^{-1}$.  The energy gap 
is to a remarkable extent determined by the available $^1S_0$ phase shifts.  
In the same figure we also report the results (dot-dashed line) 
obtained using the effective range approximation to the phase shifts: 
$$
       k\cot \delta(k)=-{{1}\over{a_0}}+{{1}\over{2}}r_0 k^2, \eqno(6)
$$
where $a_0=-18.8\pm 0.3$ fm and $r_0=2.75\pm 0.11$ fm are the singlet 
neutron-neutron scattering length and effective range, respectively.  
In this case an analytic expression can be obtained for $\lambda v^2(k)$, as 
shown in Ref.\ [16]:
$$
  \lambda v^2(k)=-{{1}\over{\sqrt{k^2+{{r_0^2}\over{4}}(k^2+\alpha^2)^2}}}
                \sqrt{{k^2+\beta_2^2}\over{k^2-\beta_1^2}},\eqno(7)
$$
with $\alpha^2=-2/ar_0$, $\beta_1\approx-0.0498\;{\rm fm}^{-1}$, 
and $\beta_2\approx 0.777\;{\rm fm}^{-1}$.  
The phase shifts using this approximation are positive at all energies, 
and this is reflected in Eq.\ (7) where $\lambda v^2(k)$ 
is attractive for all $k$.  From Fig.\ 1 we see that 
below $k_F=0.5\;{\rm fm}^{-1}$ the energy gap can with reasonable 
accuracy be calculated with the interaction obtained directly from 
the effective range approximation.  
One can therefore say that 
at densities below $k_F=0.5\;{\rm fm}^{-1}$, and at the crudest level 
of sophistication in many-body theory,  the superfluid properties 
of neutron matter are determined by just two parameters, namely 
the free-space scattering length and effective range. At such densities,
more complicated many-body terms are also less important.
Also interesting is the fact that the phase shifts predict the position 
of the first zero of $\Delta(k)$ in momentum space, since we see from 
Eq.\ (4) that $\Delta(k)=\Delta_F v(k)=0$ first for $\delta(k)=0$, 
which occurs at $E_{\rm lab}\approx 248.5$ MeV (pp scattering) 
corresponding to $k\approx 
1.74\;{\rm fm}^{-1}$.  This is in good agreement with the results of 
Khodel et al.\ [8].  In Ref.\ [8] it is 
also shown that this first zero of the gap function determines the 
Fermi momentum at which $\Delta_F=0$.  Our results therefore indicate 
that this Fermi momentum is in fact given by the energy at which 
the $^1S_0$ phase shifts become negative. 

Thus, the quantitative features 
of $^1S_0$ pairing in neutron matter can be obtained directly from 
the $^1S_0$ phase shifts. This happens because the NN interaction 
is very nearly rank-one separable in this channel due to the presence 
of a bound state at zero energy.  This explains why all bare NN interactions 
give nearly identical results for the $^1S_0$ energy gap in lowest-order 
BCS calculations.  
%\vfill\eject
%\topinsert
%\vskip 6.1 truecm
%\endinsert
%\noindent
%{\bf Figure 2.} $^1S_0$ energy gap in nuclear matter calculated with  
%                 the CD-Bonn potential  
%                 compared with the direct calculation from the 
%                 $^1S_0$ np and pp phase shifts.  Also shown are the results 
%     for neutron matter with the CD-Bonn potential.
%\vskip 28 truept
%\noindent
However, 
it should be mentioned that this agreement 
is not likely to survive in a more refined calculation, for instance 
if one includes the density and spin-density fluctuations in the 
effective pairing interaction like in e.g., Refs.\ [10,11].  
Other partial waves will then be involved, and the simple arguments 
employed here will, of course, no longer apply.  Our reasoning here
applies also only
to a partial wave where the $T$-matrix (almost) has a pole, and we have
neglected the fact that the phase shifts become negative at higher
energies.  
\begin{figure}
% GNUPLOT: LaTeX picture with Postscript
\setlength{\unitlength}{0.1bp}
\special{!
%!PS-Adobe-2.0
%%Creator: gnuplot
%%DocumentFonts: Helvetica
%%BoundingBox: 50 50 770 554
%%Pages: (atend)
%%EndComments
/gnudict 40 dict def
gnudict begin
/Color false def
/Solid false def
/gnulinewidth 5.000 def
/vshift -33 def
/dl {10 mul} def
/hpt 31.5 def
/vpt 31.5 def
/M {moveto} bind def
/L {lineto} bind def
/R {rmoveto} bind def
/V {rlineto} bind def
/vpt2 vpt 2 mul def
/hpt2 hpt 2 mul def
/Lshow { currentpoint stroke M
  0 vshift R show } def
/Rshow { currentpoint stroke M
  dup stringwidth pop neg vshift R show } def
/Cshow { currentpoint stroke M
  dup stringwidth pop -2 div vshift R show } def
/DL { Color {setrgbcolor Solid {pop []} if 0 setdash }
 {pop pop pop Solid {pop []} if 0 setdash} ifelse } def
/BL { stroke gnulinewidth 2 mul setlinewidth } def
/AL { stroke gnulinewidth 2 div setlinewidth } def
/PL { stroke gnulinewidth setlinewidth } def
/LTb { BL [] 0 0 0 DL } def
/LTa { AL [1 dl 2 dl] 0 setdash 0 0 0 setrgbcolor } def
/LT0 { PL [] 0 1 0 DL } def
/LT1 { PL [4 dl 2 dl] 0 0 1 DL } def
/LT2 { PL [2 dl 3 dl] 1 0 0 DL } def
/LT3 { PL [1 dl 1.5 dl] 1 0 1 DL } def
/LT4 { PL [5 dl 2 dl 1 dl 2 dl] 0 1 1 DL } def
/LT5 { PL [4 dl 3 dl 1 dl 3 dl] 1 1 0 DL } def
/LT6 { PL [2 dl 2 dl 2 dl 4 dl] 0 0 0 DL } def
/LT7 { PL [2 dl 2 dl 2 dl 2 dl 2 dl 4 dl] 1 0.3 0 DL } def
/LT8 { PL [2 dl 2 dl 2 dl 2 dl 2 dl 2 dl 2 dl 4 dl] 0.5 0.5 0.5 DL } def
/P { stroke [] 0 setdash
  currentlinewidth 2 div sub M
  0 currentlinewidth V stroke } def
/D { stroke [] 0 setdash 2 copy vpt add M
  hpt neg vpt neg V hpt vpt neg V
  hpt vpt V hpt neg vpt V closepath stroke
  P } def
/A { stroke [] 0 setdash vpt sub M 0 vpt2 V
  currentpoint stroke M
  hpt neg vpt neg R hpt2 0 V stroke
  } def
/B { stroke [] 0 setdash 2 copy exch hpt sub exch vpt add M
  0 vpt2 neg V hpt2 0 V 0 vpt2 V
  hpt2 neg 0 V closepath stroke
  P } def
/C { stroke [] 0 setdash exch hpt sub exch vpt add M
  hpt2 vpt2 neg V currentpoint stroke M
  hpt2 neg 0 R hpt2 vpt2 V stroke } def
/T { stroke [] 0 setdash 2 copy vpt 1.12 mul add M
  hpt neg vpt -1.62 mul V
  hpt 2 mul 0 V
  hpt neg vpt 1.62 mul V closepath stroke
  P  } def
/S { 2 copy A C} def
end
}
\begin{picture}(3600,2160)(0,0)
\special{"
gnudict begin
gsave
50 50 translate
0.100 0.100 scale
0 setgray
/Helvetica findfont 100 scalefont setfont
newpath
-500.000000 -500.000000 translate
LTa
600 251 M
2817 0 V
600 251 M
0 1858 V
LTb
600 251 M
63 0 V
2754 0 R
-63 0 V
600 437 M
63 0 V
2754 0 R
-63 0 V
600 623 M
63 0 V
2754 0 R
-63 0 V
600 808 M
63 0 V
2754 0 R
-63 0 V
600 994 M
63 0 V
2754 0 R
-63 0 V
600 1180 M
63 0 V
2754 0 R
-63 0 V
600 1366 M
63 0 V
2754 0 R
-63 0 V
600 1552 M
63 0 V
2754 0 R
-63 0 V
600 1737 M
63 0 V
2754 0 R
-63 0 V
600 1923 M
63 0 V
2754 0 R
-63 0 V
600 2109 M
63 0 V
2754 0 R
-63 0 V
600 251 M
0 63 V
0 1795 R
0 -63 V
976 251 M
0 63 V
0 1795 R
0 -63 V
1351 251 M
0 63 V
0 1795 R
0 -63 V
1727 251 M
0 63 V
0 1795 R
0 -63 V
2102 251 M
0 63 V
0 1795 R
0 -63 V
2478 251 M
0 63 V
0 1795 R
0 -63 V
2854 251 M
0 63 V
0 1795 R
0 -63 V
3229 251 M
0 63 V
0 1795 R
0 -63 V
600 251 M
2817 0 V
0 1858 V
-2817 0 V
600 251 L
LT0
3114 1946 M
180 0 V
788 278 M
2 1 V
2 1 V
4 2 V
5 2 V
6 3 V
8 3 V
8 4 V
10 5 V
11 6 V
12 7 V
13 8 V
14 9 V
16 10 V
16 12 V
18 13 V
18 14 V
20 16 V
21 18 V
21 19 V
23 21 V
23 22 V
25 24 V
25 26 V
26 27 V
27 29 V
28 30 V
29 32 V
29 33 V
31 34 V
31 34 V
31 36 V
33 37 V
32 37 V
34 38 V
34 37 V
35 38 V
35 37 V
35 37 V
36 36 V
36 35 V
37 33 V
37 33 V
37 31 V
38 29 V
37 27 V
38 25 V
38 22 V
38 21 V
38 18 V
39 15 V
38 13 V
38 9 V
38 6 V
38 2 V
37 0 V
38 -2 V
37 -5 V
37 -6 V
37 -8 V
36 -11 V
36 -13 V
35 -15 V
35 -22 V
35 -26 V
34 -29 V
34 -32 V
32 -33 V
33 -35 V
31 -37 V
31 -37 V
31 -38 V
29 -39 V
29 -38 V
28 -38 V
27 -38 V
26 -38 V
25 -36 V
25 -36 V
23 -34 V
23 -33 V
21 -32 V
21 -31 V
20 -29 V
18 -27 V
18 -26 V
16 -24 V
16 -22 V
14 -20 V
13 -19 V
12 -18 V
11 -15 V
10 -14 V
8 -12 V
8 -10 V
6 -9 V
5 -7 V
4 -5 V
2 -4 V
2 -2 V
LT1
3114 1846 M
180 0 V
788 282 M
2 1 V
2 1 V
4 2 V
5 3 V
6 3 V
8 3 V
8 5 V
10 5 V
11 7 V
12 7 V
13 8 V
14 10 V
16 11 V
16 12 V
18 13 V
18 15 V
20 16 V
21 18 V
21 19 V
23 22 V
23 22 V
25 25 V
25 26 V
26 27 V
27 29 V
28 30 V
29 32 V
29 32 V
31 34 V
31 35 V
31 36 V
33 36 V
32 36 V
34 37 V
34 37 V
35 37 V
35 37 V
35 36 V
36 35 V
36 34 V
37 33 V
37 32 V
37 30 V
38 28 V
37 26 V
38 24 V
38 22 V
38 19 V
38 17 V
39 13 V
38 12 V
38 8 V
38 4 V
38 2 V
37 -2 V
38 -4 V
37 -8 V
37 -10 V
37 -14 V
36 -17 V
36 -19 V
35 -22 V
35 -25 V
35 -27 V
34 -28 V
34 -31 V
32 -32 V
33 -34 V
31 -36 V
31 -36 V
31 -37 V
29 -37 V
29 -36 V
28 -37 V
27 -37 V
26 -36 V
25 -35 V
25 -34 V
23 -32 V
23 -31 V
21 -29 V
21 -28 V
20 -26 V
18 -24 V
18 -22 V
16 -21 V
16 -19 V
14 -17 V
13 -15 V
12 -14 V
11 -12 V
10 -10 V
8 -10 V
8 -7 V
6 -7 V
5 -5 V
4 -4 V
2 -3 V
2 -1 V
LT4
3114 1746 M
180 0 V
788 327 M
2 0 V
2 0 V
4 0 V
5 -1 V
6 0 V
8 0 V
8 1 V
10 1 V
11 1 V
12 2 V
13 2 V
14 4 V
16 4 V
16 5 V
18 7 V
18 8 V
20 10 V
21 12 V
21 13 V
23 15 V
23 17 V
25 20 V
25 21 V
26 24 V
27 26 V
28 30 V
29 32 V
29 33 V
31 36 V
31 37 V
31 39 V
33 38 V
32 38 V
34 38 V
34 39 V
35 38 V
35 37 V
35 35 V
36 35 V
36 34 V
37 33 V
37 32 V
37 34 V
38 33 V
37 32 V
38 30 V
38 28 V
38 24 V
38 22 V
39 19 V
38 16 V
38 13 V
38 10 V
38 6 V
37 4 V
38 0 V
37 -3 V
37 -8 V
37 -11 V
36 -13 V
36 -17 V
35 -19 V
35 -18 V
35 -21 V
34 -23 V
34 -28 V
32 -31 V
33 -39 V
31 -50 V
31 -53 V
31 -53 V
29 -53 V
29 -51 V
28 -49 V
27 -40 V
26 -36 V
25 -35 V
25 -32 V
23 -31 V
23 -29 V
21 -28 V
21 -26 V
20 -25 V
18 -23 V
18 -22 V
16 -21 V
16 -20 V
14 -18 V
13 -17 V
12 -16 V
11 -14 V
10 -13 V
8 -11 V
8 -10 V
6 -9 V
5 -7 V
4 -5 V
2 -3 V
2 -3 V
stroke
grestore
end
showpage
}
\put(3054,1746){\makebox(0,0)[r]{Phase Shifts}}
\put(3054,1846){\makebox(0,0)[r]{CD-Bonn, Neutron Matter}}
\put(3054,1946){\makebox(0,0)[r]{CD-Bonn Nuclear Matter}}
\put(2008,51){\makebox(0,0){$k_F$ (fm$^{-1}$)}}
\put(100,1180){%
\special{ps: gsave currentpoint currentpoint translate
270 rotate neg exch neg exch translate}%
\makebox(0,0)[b]{\shortstack{Pairing gap $\Delta (k_F)$ (MeV)}}%
\special{ps: currentpoint grestore moveto}%
}
\put(3229,151){\makebox(0,0){1.4}}
\put(2854,151){\makebox(0,0){1.2}}
\put(2478,151){\makebox(0,0){1}}
\put(2102,151){\makebox(0,0){0.8}}
\put(1727,151){\makebox(0,0){0.6}}
\put(1351,151){\makebox(0,0){0.4}}
\put(976,151){\makebox(0,0){0.2}}
\put(600,151){\makebox(0,0){0}}
\put(540,2109){\makebox(0,0)[r]{5}}
\put(540,1923){\makebox(0,0)[r]{4.5}}
\put(540,1737){\makebox(0,0)[r]{4}}
\put(540,1552){\makebox(0,0)[r]{3.5}}
\put(540,1366){\makebox(0,0)[r]{3}}
\put(540,1180){\makebox(0,0)[r]{2.5}}
\put(540,994){\makebox(0,0)[r]{2}}
\put(540,808){\makebox(0,0)[r]{1.5}}
\put(540,623){\makebox(0,0)[r]{1}}
\put(540,437){\makebox(0,0)[r]{0.5}}
\put(540,251){\makebox(0,0)[r]{0}}
\end{picture}
	\caption{$^1S_0$ energy gap in nuclear matter calculated with  
                 the CD-Bonn potential  
                 compared with the direct calculation from the 
                 $^1S_0$ np and pp phase shifts.  Also shown are the results 
     for neutron matter with the CD-Bonn potential.}
\end{figure}
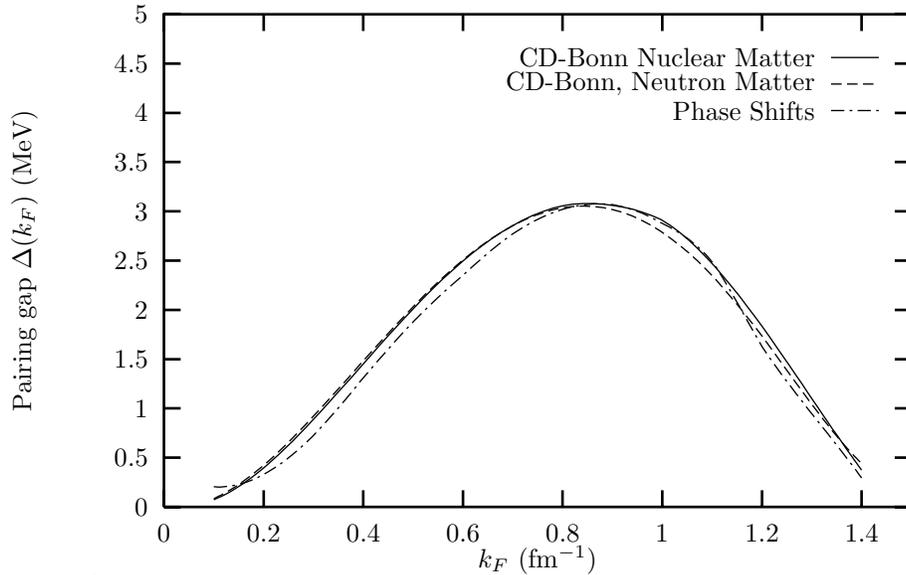	
The calculation of the $^1S_0$ gap in symmetric nuclear matter is  
closely related to the one for neutron matter.  In fact, with 
charge-independent forces, like the older Bonn potentials, and 
free single-particle energies one would, of course, obtain 
exactly the same results.  However, the new potentials 
on the market are charge-dependent, in order to achieve high quality fits  
to both np and pp scattering data, and therefore we should 
in principle  solve three 
coupled gap equations for neutron-neutron (nn), proton-proton (pp), 
and neutron-proton (np) pairing:
$$
   \Delta_i(k)=-{{1}\over{\pi}}\int_0^{\infty}dk'k'^2 V_i(k,k')
   {{\Delta_i(k')}\over{E(k')}}, \eqno(8)
$$
where $i$=nn, pp and np, and the quasiparticle energy is still given by 
$E(k)=\sqrt{(\epsilon(k)-\epsilon(k_F))^2+\Delta(k)^2}$, but the energy 
gap is now given by 
$$
  \Delta(k)^2=
  \Delta_{\rm nn}(k)^2+\Delta_{\rm pp}(k)^2+\Delta_{\rm np}(k)^2. \eqno(9)
$$
Solving these equations, both with the CD-Bonn potential and 
with the phase shift approximations we get the results shown in Fig.\ 2.
For comparison we have in the same figure plotted the 
results for pure neutron matter with the CD-Bonn potential (dashed line). 
From the figure it is clear that the phase shift approximation works well 
also in this case.   As could be expected, the results 
are very close to those obtained earlier with charge-independent 
interactions [7].
%\vskip 48 truept

%\centerline{\bf 3. FEATURES OF PAIRING CORRELATIONS IN INFINITE MATTER}
%\vskip 12 truept

\section{FEATURES OF PAIRING CORRELATIONS IN INFINITE MATTER}

An important  length scale of e.g., a neutron superfluid   
is the coherence length. From a microscopic point of view the 
coherence length represents 
the squared mean distance   of two paired particles 
(a Cooper pair of neutrons)  
on top of the Fermi surface. 
The magnitude of this quantity affects several of the physical properties 
of a neutron star crust. 
First of all, neutrons paired in a singlet state form quantized vortices 
induced by the rotational 
state of the star. These   can pin to the nuclei present in the 
crust, possibly leading to the observed sudden release of angular momentum 
known as pulsar glitches. The magnitude of the pinning force 
depends on the size of the vortex cores, 
which is equal to the coherence length of the neutron superfluid. A second 
question is how properties of the neutron 
superfluid change due to  the inhomogeneous 
environment of a neutron star crust, a problem related to the average 
thermodynamical properties of neutron  matter [1].  
The typical dimension of nuclei 
in the inner crust of a neutron star 
is $R_{N}\approx 4 - 6\;{\rm fm}$.  This number is, in an appropriate range 
of densities, comparable to the coherence length $\xi$ as 
estimated from existing BCS calculations.
Clearly, the coherence length represents a critical parameter by which
one can 
establish the behavior of an inhomogeneous superfluid. 
It sets the scale for the possible spatial variation of the pairing 
properties of the system, and thus plays a role if some 
inhomogeneities are present in the system at a length scale 
comparable to it. 

The coherence length can be easily evaluated from the wavefunction 
$\phi({\bf r})$ of the 
relative motion of the two neutrons in a Cooper pair, ${\bf r}$ 
being the relative coordinate of the two particles.   
The coherence length $\xi$ is given by 
$$
 \xi^2={{\int d^3r|\phi({\bf r})|^2r^2}\over{\int d^3r|\phi({\bf r})|^2}}={{\int_0^{\infty}dkk^2|\partial\chi(k)/\partial k|^2}\over{\int_{0}^{\infty}dkk^2|\chi(k)|^2}}, \eqno(10)
$$
with
$\chi({\bf k})$ being the wavefunction of a Cooper pair 
in momentum space.
This equation is particularly suited for numerical computation,  
since the  BCS equations for a uniform system are solved in momentum space,
as discussed in the previous section. 
The wavefunction of the Cooper pair in momentum space is 
given by (apart from an unimportant normalization constant)  
$$
     \chi({ k})={\Delta(k) \over E(k)}, \eqno(11)
$$ 
where $\Delta(k)$ is the $k$-dependent pairing gap, 
while $E(k)$ is the energy denominator in Eq.\ (1).  

As mentioned in the introduction, 
the NN potential in the $^1S_0$ channel yields actually  
two so-called weak-coupling limits. This happens when the potential is weak 
and attractive for large interparticle
spacings and when the potential becomes repulsive at $r=0.57$ fm. 
This corresponds to the Fermi momentum $k_F=1.74$ fm$^{-1}$ discussed
in the previous section in connection with the phase shift analyses.
The other weak-coupling limit, i.e., when $r$ is large and the potential
tends to zero, corresponds to small values of the Fermi momentum.
The pairing gap decays exponentially to zero in the low-density
limit [8]. 
What can be thought of as a 
strong coupling limit takes place where the NN potential
reaches its maximum, at roughly $\sim 1$ fm, see e.g., Ref.\ [15] for 
a discussion of various features of the NN potential. The pairing
gaps in Figs.\ 1 and 2 have their maxima at the density 
which corresponds roughly to the maximum of the NN interaction in the  
$^1S_0$ channel.

In the weak-coupling limits, we have that the bound pairs overlap in $r$-space,
or stated differently, 
that the wave function in Eq.\ (11) is strongly peaked in momentum
space at the value of the corresponding Fermi surface. 
The coherence length $\xi$ is in this case much larger than the
typical interparticle spacing, see
e.g., the discussion in Ref.\ [12]. 
As stated in the introduction, we will with weak-coupling mean
that the gas of Cooper pairs will be surrounded by unpaired fermions and
the typical coherence length is large compared with the interparticle
spacing, and the bound pairs overlap. 

If we have a strong coupling limit, the Cooper pairs at the Fermi surface
have only small 
overlaps, a fact which means in turn that the pair wave function in Eq.\ (11)
extends further out in $k$-space, or that the Cooper pair 
is more localized in $r$-space. The coherence length should then be small, 
of the order or smaller than the interparticle spacing.
 
The wave function of Eq.\ (11) 
for the various possible coupling regimes
is shown in Fig.\ 3. There we plot $\chi({ k})$ for five values
of $k_F$, 0.03, 04, 0.8, 1.2 and 1.4 fm$^{-1}$, employing the pairing gap
from the previous section obtained with the CD-Bonn potential. The phase-shift
approximation or the Nijmegen potentials yield essentially the same results.
 
Clearly, at low values, $k_F=0.03$ fm$^{-1}$ in Fig.\ 3, 
of the Fermi momentum, corresponding  
to one of the weak-coupling regimes, the wave function is
strongly peaked in momentum space. Similarly, for densities where the 
NN interaction changes from being attractive to repulsive, we have the other
weak-coupling regime. The qualitative form of the wave function
at $k_F=1.4$ fm$^{-1}$ resembles much that at low densities. 
For densities corresponding to $r$-values where the potential is close
to its maximum, $k_F\sim 0.7-1.2$ fm$^{-1}$, one sees that the wave function
in $k$-space is much more spread out, possibly implying that the coherence length is
smaller and that the Cooper pairs have only small overlaps.

The fact that the NN interaction in the  $^1S_0$ channel is large and
attractive at certain values of $k_F$ (up to five times larger than the
Fermi energy) and that the wave functions in Fig.\ 3 extend over several values
of $k$, may lead one to conclude that one could speak of 
bound fermion pairs which 
can be treated as a gas of point bosons. One expects then the system
to undergo a Bose-Einstein condensation into a single quantum
state with total momentum $k=0$, as discussed in depth in Ref.\ [12].
However, such a conclusion for singlet pairing in neutron or nuclear
matter is wrong. If one calculates the coherence length using Eq.\ (10)
for the above Fermi momenta, one finds that for all Fermi momenta
the coherence length is much larger than the typical interparticle
spacing. Eq.\ (10) gives 
$\xi=$ 388.0, 4.8, 5.2, 13.2 and 53.5 fm for
$k_F=$ 0.03, 04, 0.8, 1.2 and 1.4 fm$^{-1}$, respectively. Even the smallest
values are of the size of the radius of nuclei found in the crust
of a neutron star. 
If one also observes that screening effects yield even larger
coherence lengths, see e.g., [19], one can conclude that for
singlet pairing in neutron or nuclear matter,  
the gas of Cooper pairs has  
a typical coherence length which is large compared with the interparticle
spacing, the Cooper pairs overlap and fermion exchange may become dominant.
 
From Fig.\ 3 one also notices that for the chosen Fermi momenta, the pair 
wave function does not vanish at $k=0$.
This behavior is easy to understand if we again employ a rank-one separable
interaction. Eq.\ (11) reads then
$$
     \chi({ k})={{v(k)\Delta(k_F)} \over 
     {\sqrt{(\epsilon(k)-\epsilon(k_F))^2+v(k)^2\Delta(k_F)^2}}}, \eqno(12)
$$ 
which at $k=0$ simplifies to 
$$
     \chi({ 0})={{v(0)\Delta(k_F)} \over 
     {\sqrt{\epsilon(k_F)^2+v(0)^2\Delta(k_F)^2}}}, \eqno(13)
$$
where $\lambda$ of Eq.\ (2) is set equal to one. The $v(0)$ part of the 
potential can in turn be determined directly from the scattering matrix
at $k=0$. In that limit the scattering matrix equals $-a_0$, where $a_0$ is the 
scattering length, $a_0=-18.8\pm 0.3$ fm for the neutron-neutron potential.
For a rank-one separable potential, the on-shell scattering matrix at $k=0$ is
given by [15]
$$
      T(k=0)=-a_0={{v(0)^2}\over 
      {1+ {{2}\over{\pi}} \int_{0}^{\infty}dq v(q)^2}}. \eqno(14)
$$ 
If the $^1S_0$ channel really has a bound state at $k=0$, the denominator
should diverge, which in turn means that the scattering 
length should be  $a_0=-\infty$. The fact that the scattering length is
finite implies that $v(0)$ is finite.
Eq.\ (13) can be rewritten as
$$
     {{\chi({ 0})^2\epsilon(k_F)^2}\over{\Delta(k_F)^2(1-\chi({ 0})^2)}}=
     v(0)^2. \eqno(15)
$$ 
%\vfill\eject
%\topinsert
%\vskip 6.1 truecm
%\endinsert
%\noindent
%{\bf Figure 3.} Wavefunctions in momentum space, $\chi (k)$ 
%         for five different values of the Fermi momentum, 
%         $k_F=0.03$ fm$^{-1}$, $k_F=0.4$ fm$^{-1}$, $k_F=0.8$ fm$^{-1}$, 
%         $k_F=1.2$ fm$^{-1}$ and $k_F=1.4$ fm$^{-1}$.
%\vskip 28 truept
%\noindent
When $k_F \rightarrow 0$, the gap behaves asymptotically as [8]
$$
   \Delta(k_F)\sim 8\epsilon(k_F)e^{-1/\gamma -2}, \eqno(16)
$$
where $\gamma = -2k_f a_0/\pi$. Inserting Eq.\ (16) into Eq.\ (15) 
then yields that the wave function $\chi({ 0}) \rightarrow 0$
when  $k_F \rightarrow 0$.
In the other weak-coupling limit when $k_F=1.74$ fm$^{-1}$, 
i.e., where the potential changes sign, $v(k_F)=0$, which inserted in 
Eq.\ (13) shows that the wave function goes to 
$\chi({ 0}) \rightarrow 0$
when  $k_F \rightarrow 1.74$ fm$^{-1}$.
\begin{figure}
% GNUPLOT: LaTeX picture with Postscript
\setlength{\unitlength}{0.1bp}
\special{!
%!PS-Adobe-2.0
%%Creator: gnuplot
%%DocumentFonts: Helvetica
%%BoundingBox: 50 50 770 554
%%Pages: (atend)
%%EndComments
/gnudict 40 dict def
gnudict begin
/Color false def
/Solid false def
/gnulinewidth 5.000 def
/vshift -33 def
/dl {10 mul} def
/hpt 31.5 def
/vpt 31.5 def
/M {moveto} bind def
/L {lineto} bind def
/R {rmoveto} bind def
/V {rlineto} bind def
/vpt2 vpt 2 mul def
/hpt2 hpt 2 mul def
/Lshow { currentpoint stroke M
  0 vshift R show } def
/Rshow { currentpoint stroke M
  dup stringwidth pop neg vshift R show } def
/Cshow { currentpoint stroke M
  dup stringwidth pop -2 div vshift R show } def
/DL { Color {setrgbcolor Solid {pop []} if 0 setdash }
 {pop pop pop Solid {pop []} if 0 setdash} ifelse } def
/BL { stroke gnulinewidth 2 mul setlinewidth } def
/AL { stroke gnulinewidth 2 div setlinewidth } def
/PL { stroke gnulinewidth setlinewidth } def
/LTb { BL [] 0 0 0 DL } def
/LTa { AL [1 dl 2 dl] 0 setdash 0 0 0 setrgbcolor } def
/LT0 { PL [] 0 1 0 DL } def
/LT1 { PL [4 dl 2 dl] 0 0 1 DL } def
/LT2 { PL [2 dl 3 dl] 1 0 0 DL } def
/LT3 { PL [1 dl 1.5 dl] 1 0 1 DL } def
/LT4 { PL [5 dl 2 dl 1 dl 2 dl] 0 1 1 DL } def
/LT5 { PL [4 dl 3 dl 1 dl 3 dl] 1 1 0 DL } def
/LT6 { PL [2 dl 2 dl 2 dl 4 dl] 0 0 0 DL } def
/LT7 { PL [2 dl 2 dl 2 dl 2 dl 2 dl 4 dl] 1 0.3 0 DL } def
/LT8 { PL [2 dl 2 dl 2 dl 2 dl 2 dl 2 dl 2 dl 4 dl] 0.5 0.5 0.5 DL } def
/P { stroke [] 0 setdash
  currentlinewidth 2 div sub M
  0 currentlinewidth V stroke } def
/D { stroke [] 0 setdash 2 copy vpt add M
  hpt neg vpt neg V hpt vpt neg V
  hpt vpt V hpt neg vpt V closepath stroke
  P } def
/A { stroke [] 0 setdash vpt sub M 0 vpt2 V
  currentpoint stroke M
  hpt neg vpt neg R hpt2 0 V stroke
  } def
/B { stroke [] 0 setdash 2 copy exch hpt sub exch vpt add M
  0 vpt2 neg V hpt2 0 V 0 vpt2 V
  hpt2 neg 0 V closepath stroke
  P } def
/C { stroke [] 0 setdash exch hpt sub exch vpt add M
  hpt2 vpt2 neg V currentpoint stroke M
  hpt2 neg 0 R hpt2 vpt2 V stroke } def
/T { stroke [] 0 setdash 2 copy vpt 1.12 mul add M
  hpt neg vpt -1.62 mul V
  hpt 2 mul 0 V
  hpt neg vpt 1.62 mul V closepath stroke
  P  } def
/S { 2 copy A C} def
end
}
\begin{picture}(3600,2160)(0,0)
\special{"
gnudict begin
gsave
50 50 translate
0.100 0.100 scale
0 setgray
/Helvetica findfont 100 scalefont setfont
newpath
-500.000000 -500.000000 translate
LTa
600 251 M
2817 0 V
600 251 M
0 1858 V
LTb
600 251 M
63 0 V
2754 0 R
-63 0 V
600 561 M
63 0 V
2754 0 R
-63 0 V
600 870 M
63 0 V
2754 0 R
-63 0 V
600 1180 M
63 0 V
2754 0 R
-63 0 V
600 1490 M
63 0 V
2754 0 R
-63 0 V
600 1799 M
63 0 V
2754 0 R
-63 0 V
600 2109 M
63 0 V
2754 0 R
-63 0 V
600 251 M
0 63 V
0 1795 R
0 -63 V
1163 251 M
0 63 V
0 1795 R
0 -63 V
1727 251 M
0 63 V
0 1795 R
0 -63 V
2290 251 M
0 63 V
0 1795 R
0 -63 V
2854 251 M
0 63 V
0 1795 R
0 -63 V
3417 251 M
0 63 V
0 1795 R
0 -63 V
600 251 M
2817 0 V
0 1858 V
-2817 0 V
600 251 L
LT0
3114 1946 M
180 0 V
600 357 M
1 0 V
1 0 V
1 1 V
1 0 V
1 1 V
1 1 V
1 2 V
2 2 V
1 3 V
2 4 V
1 5 V
2 7 V
2 10 V
2 13 V
2 19 V
2 28 V
3 45 V
2 83 V
2 197 V
3 724 V
1 297 V
2 -700 V
2 -474 V
3 -147 V
3 -69 V
2 -39 V
3 -25 V
3 -18 V
3 -12 V
3 -10 V
3 -7 V
3 -6 V
3 -5 V
3 -4 V
3 -3 V
3 -3 V
3 -3 V
3 -2 V
3 -1 V
3 -2 V
3 -1 V
3 -1 V
3 -1 V
3 -1 V
3 -1 V
3 -1 V
2 -1 V
3 0 V
3 -1 V
2 0 V
2 -1 V
3 0 V
2 0 V
2 -1 V
2 0 V
2 0 V
2 -1 V
2 0 V
2 0 V
1 0 V
2 0 V
1 0 V
2 -1 V
1 0 V
1 0 V
1 0 V
1 0 V
1 0 V
1 0 V
4 0 V
6 -1 V
9 0 V
12 -1 V
15 -1 V
17 0 V
20 -1 V
23 0 V
25 0 V
28 -1 V
30 0 V
33 0 V
35 0 V
38 0 V
40 -1 V
42 0 V
44 0 V
46 0 V
48 0 V
51 0 V
52 0 V
53 0 V
56 0 V
57 0 V
58 0 V
60 0 V
61 0 V
62 0 V
63 0 V
64 0 V
65 0 V
65 0 V
67 0 V
66 0 V
67 0 V
67 0 V
67 0 V
68 0 V
67 0 V
67 0 V
66 0 V
66 0 V
66 0 V
65 0 V
LT1
3114 1846 M
180 0 V
600 932 M
1 0 V
1 0 V
1 0 V
2 0 V
3 0 V
3 0 V
4 0 V
3 0 V
5 1 V
5 0 V
5 1 V
6 1 V
6 2 V
6 1 V
7 2 V
8 2 V
7 3 V
8 3 V
9 4 V
9 4 V
9 5 V
9 6 V
10 6 V
10 8 V
10 8 V
10 9 V
11 11 V
11 11 V
11 13 V
11 14 V
11 16 V
12 18 V
11 19 V
12 21 V
11 24 V
12 26 V
12 28 V
11 32 V
12 34 V
12 37 V
11 40 V
12 44 V
11 46 V
12 49 V
11 51 V
11 51 V
11 51 V
10 48 V
11 43 V
10 35 V
10 25 V
9 13 V
10 -1 V
9 -13 V
9 -25 V
8 -35 V
8 -42 V
8 -47 V
7 -48 V
7 -48 V
7 -47 V
6 -44 V
5 -42 V
6 -37 V
5 -34 V
4 -31 V
4 -27 V
3 -23 V
3 -20 V
3 -16 V
2 -13 V
2 -10 V
1 -7 V
0 -3 V
1 -6 V
3 -18 V
6 -32 V
8 -45 V
10 -54 V
13 -60 V
15 -64 V
17 -65 V
20 -64 V
22 -59 V
24 -56 V
26 -50 V
29 -45 V
31 -41 V
32 -35 V
35 -31 V
37 -28 V
38 -24 V
41 -20 V
42 -19 V
44 -15 V
45 -14 V
47 -12 V
48 -11 V
50 -9 V
50 -8 V
52 -7 V
54 -6 V
54 -5 V
55 -5 V
55 -4 V
57 -3 V
57 -4 V
58 -2 V
58 -3 V
58 -2 V
58 -1 V
59 -2 V
58 -1 V
59 -1 V
58 -1 V
58 -1 V
58 -1 V
LT2
3114 1746 M
180 0 V
600 709 M
1 0 V
2 0 V
3 0 V
4 0 V
4 0 V
6 0 V
6 0 V
7 0 V
8 0 V
9 1 V
9 0 V
11 0 V
11 1 V
12 1 V
12 1 V
13 1 V
14 1 V
15 2 V
15 2 V
16 2 V
16 2 V
17 3 V
17 3 V
18 4 V
18 4 V
19 4 V
19 5 V
20 6 V
19 6 V
20 7 V
21 8 V
20 9 V
21 10 V
21 10 V
21 12 V
21 14 V
21 14 V
21 17 V
21 18 V
21 20 V
21 23 V
21 25 V
20 29 V
21 31 V
20 36 V
20 40 V
19 45 V
19 50 V
19 56 V
18 62 V
18 68 V
17 73 V
17 77 V
17 78 V
15 74 V
15 63 V
15 47 V
14 23 V
13 -2 V
13 -26 V
11 -44 V
11 -57 V
11 -62 V
9 -63 V
9 -60 V
8 -54 V
7 -49 V
6 -43 V
6 -35 V
4 -30 V
4 -23 V
3 -18 V
2 -12 V
1 -7 V
1 -5 V
2 -15 V
5 -27 V
7 -37 V
8 -47 V
11 -53 V
13 -57 V
14 -60 V
17 -60 V
18 -59 V
21 -56 V
22 -52 V
24 -48 V
26 -44 V
27 -40 V
29 -36 V
31 -32 V
32 -28 V
34 -26 V
36 -22 V
36 -20 V
38 -18 V
40 -16 V
40 -13 V
42 -13 V
43 -11 V
43 -9 V
45 -9 V
45 -7 V
47 -7 V
47 -6 V
47 -5 V
48 -5 V
48 -4 V
49 -3 V
49 -3 V
49 -3 V
49 -2 V
49 -2 V
LT3
3114 1646 M
180 0 V
600 424 M
2 0 V
3 0 V
4 0 V
5 0 V
7 0 V
8 0 V
9 0 V
10 0 V
11 0 V
13 0 V
14 0 V
15 1 V
16 0 V
17 0 V
18 0 V
19 0 V
20 1 V
21 0 V
22 1 V
22 0 V
24 1 V
24 0 V
25 1 V
26 1 V
27 1 V
27 1 V
27 1 V
29 2 V
28 1 V
29 2 V
29 2 V
30 2 V
30 2 V
30 3 V
31 3 V
30 4 V
30 3 V
31 5 V
30 4 V
31 6 V
30 6 V
30 7 V
29 8 V
30 9 V
29 10 V
28 12 V
28 13 V
28 16 V
27 19 V
27 22 V
25 27 V
25 33 V
25 40 V
23 52 V
23 65 V
22 86 V
21 116 V
20 157 V
19 211 V
18 245 V
17 166 V
16 -63 V
15 -218 V
14 -220 V
12 -172 V
12 -125 V
10 -92 V
9 -67 V
8 -50 V
7 -37 V
5 -27 V
4 -19 V
3 -13 V
1 -7 V
1 -3 V
2 -9 V
4 -15 V
5 -19 V
8 -24 V
8 -26 V
10 -28 V
12 -27 V
14 -27 V
15 -26 V
16 -25 V
18 -22 V
20 -21 V
20 -18 V
23 -17 V
23 -15 V
25 -14 V
26 -12 V
28 -11 V
28 -10 V
30 -9 V
31 -8 V
32 -7 V
33 -6 V
33 -5 V
35 -5 V
35 -5 V
36 -4 V
37 -3 V
38 -3 V
38 -3 V
38 -2 V
34 -2 V
LT4
3114 1546 M
180 0 V
600 295 M
2 0 V
4 0 V
4 0 V
6 0 V
8 0 V
9 0 V
11 0 V
11 0 V
14 0 V
14 0 V
16 0 V
17 0 V
19 0 V
19 0 V
21 0 V
22 0 V
23 0 V
24 0 V
26 0 V
26 0 V
27 0 V
28 1 V
29 0 V
30 0 V
30 0 V
32 0 V
32 0 V
32 1 V
33 0 V
33 0 V
34 0 V
35 1 V
34 0 V
35 1 V
35 0 V
35 1 V
35 0 V
35 1 V
35 1 V
35 1 V
35 1 V
35 1 V
34 2 V
34 1 V
33 2 V
33 2 V
33 2 V
32 3 V
31 3 V
30 4 V
30 5 V
29 6 V
28 7 V
27 9 V
27 12 V
25 16 V
24 22 V
23 33 V
22 54 V
21 101 V
20 246 V
18 850 V
4 114 V
13 -743 V
16 -398 V
15 -137 V
13 -67 V
12 -39 V
10 -24 V
9 -17 V
8 -12 V
6 -8 V
5 -6 V
3 -3 V
2 -2 V
1 -1 V
1 -2 V
4 -3 V
4 -4 V
7 -6 V
7 -5 V
9 -6 V
11 -6 V
12 -6 V
13 -6 V
14 -5 V
16 -5 V
17 -4 V
19 -5 V
19 -3 V
21 -4 V
22 -3 V
23 -3 V
25 -2 V
25 -2 V
26 -2 V
27 -2 V
28 -2 V
29 -1 V
30 -1 V
31 -2 V
31 -1 V
16 0 V
stroke
grestore
end
showpage
}
\put(3054,1546){\makebox(0,0)[r]{$k_F=1.4$ fm$^{-1}$}}
\put(3054,1646){\makebox(0,0)[r]{$k_F=1.2$ fm$^{-1}$}}
\put(3054,1746){\makebox(0,0)[r]{$k_F=0.8$ fm$^{-1}$}}
\put(3054,1846){\makebox(0,0)[r]{$k_F=0.4$ fm$^{-1}$}}
\put(3054,1946){\makebox(0,0)[r]{$k_F=0.03$ fm$^{-1}$}}
\put(2008,51){\makebox(0,0){$k_F$ fm$^{-1}$}}
\put(100,1180){%
\special{ps: gsave currentpoint currentpoint translate
270 rotate neg exch neg exch translate}%
\makebox(0,0)[b]{\shortstack{$\chi (k)$}}%
\special{ps: currentpoint grestore moveto}%
}
\put(3417,151){\makebox(0,0){2.5}}
\put(2854,151){\makebox(0,0){2}}
\put(2290,151){\makebox(0,0){1.5}}
\put(1727,151){\makebox(0,0){1}}
\put(1163,151){\makebox(0,0){0.5}}
\put(600,151){\makebox(0,0){0}}
\put(540,2109){\makebox(0,0)[r]{1.2}}
\put(540,1799){\makebox(0,0)[r]{1}}
\put(540,1490){\makebox(0,0)[r]{0.8}}
\put(540,1180){\makebox(0,0)[r]{0.6}}
\put(540,870){\makebox(0,0)[r]{0.4}}
\put(540,561){\makebox(0,0)[r]{0.2}}
\put(540,251){\makebox(0,0)[r]{0}}
\end{picture}
\caption{Wavefunctions in momentum space, $\chi (k)$ 
         for five different values of the Fermi momentum, 
         $k_F=0.03$ fm$^{-1}$, $k_F=0.4$ fm$^{-1}$, $k_F=0.8$ fm$^{-1}$, 
         $k_F=1.2$ fm$^{-1}$ and $k_F=1.4$ fm$^{-1}$. }
\end{figure}
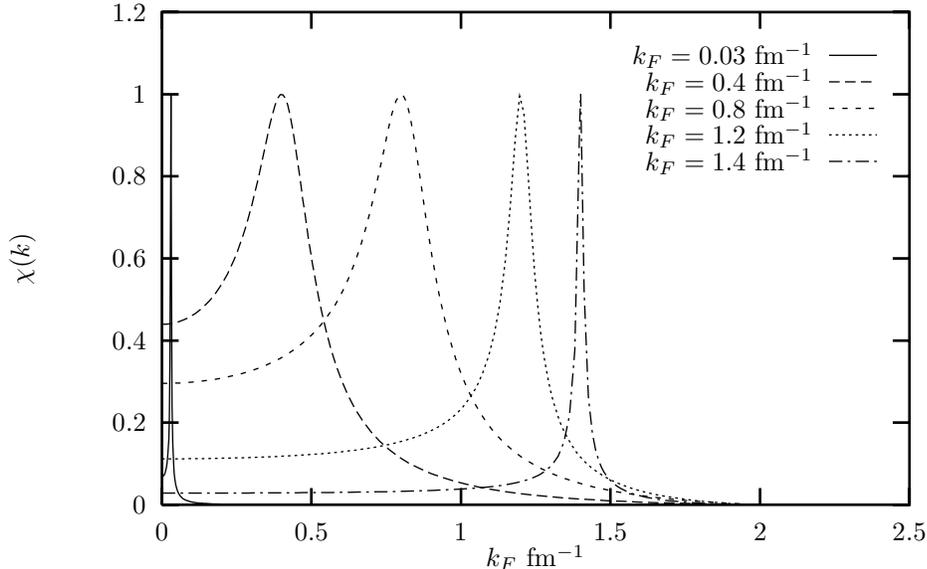
Finally, at $k=k_F$, one sees from Eq.\ (12) that $\chi({ k=k_F})=1$, 
as also seen in Fig.\ 3.

%\vskip 28 truept

%\centerline{\bf 4. CONCLUSIONS}
%\vskip 12 truept

\section{ CONCLUSIONS}

In summary, we have shown that in infinite neutron and nuclear matter, 
owing to the near rank-one separability of the NN interaction in 
the $^1S_0$ partial wave,  
we are able to compute the $^1S_0$ pairing gap directly from the NN 
phase shifts. This explains why all NN potentials which fit 
the scattering data result in almost identical $^1S_0$ pairing gaps.  
Our findings conform with the conclusions of Khodel et al. [8] 
and Carlson et al.\ [20]: 
The virtual bound state in $^1S_0$ NN scattering determines  
the features of nucleon pairing in that partial wave.    
Even though  this result    
is not likely to survive in a more refined calculation, for instance 
if one includes polarization effects in the 
effective pairing interaction like in e.g., Refs.\ [10,11], 
one can argue that our results demonstrate that upper limits for 
the value of the energy gap and for the density where a 
$^1S_0$ neutron/nucleon superfluid can exist, can be set 
directly from the $^1S_0$ phase shifts, since the polarization term 
serves to cut down the value of the gap, and leave the upper density 
for this superfluid more or less unchanged.    
These polarization terms will also enhance the already large 
coherence lengths for the singlet pairs, and one can conclude that for
singlet pairing in neutron or nuclear matter,  
the gas of Cooper pairs will be surrounded by unpaired fermions, 
the typical coherence length is large compared with the interparticle
spacing, and the bound pairs overlap.

Finally, we note that
fact that a bound state or a virtual bound state 
can be used to determine the properties of pairing in a physical system,
may be of use in studies of superfluidity and superconductivity in atomic
gases, such as
a spin-polarized $^6$Li gas, recently studied by 
Stoof et al.\  in [21].  
The scattering length of lithium
is large and negative, as is the case for the $^1S_0$ state discussed here.
Since this is a very dilute system
one can then even use an effective range
approach to the inter-particle interaction and determine
the gap uniquely for such dilute systems, by simply employing a separable
interaction of the form shown in Eq.\ (7) and discussed
in Fig.\ 1. 

We acknowledge several discussions with Marcello Baldo, Brett Carlson, 
John Clark, Manuel de Llano and Eivind Osnes. 
%\vskip 24 truept

%\centerline{\bf REFERENCES}

%\vskip 12 truept

%\end

\end{document}